\documentclass[aps,showpacs,superscriptaddress,twocolumn]{revtex4}%
\usepackage{amsfonts}
\usepackage{amsmath}
\usepackage{amssymb}
\usepackage{graphicx}%
\begin{document}
\title{ Chiral $f$-wave Topological Superfluid in Triangular Optical Lattices}
\author{Ningning Hao}
\affiliation{Beijing National Laboratory for Condensed Matter Physics and Institute of
Physics, Chinese Academy of Sciences, P. O. Box 603, Beijing 100190, China}
\affiliation{Department of Physics, Purdue University, West Lafayette, Indiana 47907, USA}
\author{Guocai Liu}
\affiliation{School of Science, Hebei University of Science and Technology, Shijiazhuang
050018, China}
\author{Ning Wu}
\affiliation{Department of Physics, Tsinghua University, Beijing 100084, China}
\author{Jiangping Hu}
\affiliation{Department of Physics, Purdue University, West Lafayette, Indiana 47907, USA}
\affiliation{Beijing National Laboratory for Condensed Matter Physics and Institute of
Physics, Chinese Academy of Sciences, P. O. Box 603, Beijing 100190, China}
\author{Yupeng Wang}
\affiliation{Beijing National Laboratory for Condensed Matter Physics and Institute of
Physics, Chinese Academy of Sciences, P. O. Box 603, Beijing 100190, China}

\begin{abstract}
We demonstrate that an exotically chiral $f$-wave topological superfluid can
be induced in cold-fermionic-atom triangular optical lattices through the
laser-field-generated effective non-Abelian gauge field, controllable Zeeman
fields and $s$-wave Feshbach resonance. We find that the chiral $f$-wave
topological superfluid is characterized by three gapless Majorana edge states
located on the boundary of the system. More interestingly, these Majorana edge
states degenerate into one Majorana fermion bound to each vortex in the
superfluid. Our proposal enlarges topological superfluid family and specifies
a unique experimentally controllable system to study the Majorana fermion physics.

\end{abstract}

\maketitle

\bigskip

\section{\bigskip Introduction}

Topological superconductors (TSCs) and superfluids (TSFs)\cite{Qi,Qi1} have
attracted considerable interest in condensed matter physics because of their
potential applications on the fault-tolerant topological quantum computation
(TQC)\cite{Nayak}. One of the remarkable features in TSCs/TSFs is the helicity
or chirality of the unconventional pairings. Unfortunately, there are very
limited natural materials\cite{Osheroff,Mackenzie} exhibiting these kinds of
unconventional pairings. Starting with the pioneer work by Fu $et$ $al$.
\cite{Fu}, recently, some new classes of hybridized systems
\cite{Sau,Mao,Qi2,Suk} have been proposed as possible candidates for TSCs,
where the unconventional pairings are induced from proximity effects of
conventional $s$-wave superconductor films. However, the impurities or
disorders in the materials hosting the electron gas increase the difficulties
to investigate the topological properties in the hybridized systems from
experiments\cite{Mourik}. Therefore, it should be not only interesting but
also necessary to design other systems that present TSC/TSF phases.

The ultra-cold atom gas associated with optical lattice technology provide an
ideal platform to realize and investigate the topological
phases\cite{Zhu,Sato,Shao,Zhang1,Zhang2} due to the controllability and
cleanity. In particular, some chiral $p$-wave TSFs\cite{Sato, Zhang1} have
been proposed based on the laser-induced artificial gauge fields
\cite{Osterloh,Ruseckas} in cold atom systems. The effect of the artificial
gauge fields is equivalent to the spin-orbit coupling, a key factor to induce
topological phase. More recently, some experimental groups reported the
realization of strong spin-orbit coupling in ultra-cold fermionic atoms gas
$^{40}$K and $^{6}$Li\cite{Wang,Cheuk}. This new technique brings the huge
hope to realize many exotic states related with spin-orbit coupling.

Recently, triangular optical lattices(TOLs) have been widely investigated in
experiments and theory, and the external fields and the different types of
interactions among the filled ultra-cold atoms can induce rich quantum phases
in the TOLs\cite{Struck,Hauke,Tieleman}. In this paper, we propose that an
exotically chiral $f$-wave TSF can be realized through the effective $k^{3}$
Rashba spin-orbit coupling(RSOC)\cite{Rashba}, Zeeman field(ZF) and $s$-wave
Feshbach resonance in triangular optical lattices(TOLs). The effective $k^{3}$
Rashba SOC and ZF are produced by the laser-atom interactions through
modulating applied laser beams. The $s$-wave Feshbach resonance is utilized to
induce the SF states\cite{Chin} of the trapped atoms. We find that there
exists a phase transition separating the TSF and normal superfluid (NSF),
which is determined by the bulk gap closing mechanism\cite{Sato3}. The TSF
resembles the SF with $f$-wave paring symmetry\cite{Hung}, which is consistent
with the geometrical symmetry of the TOLs. The chiral $f$-wave TSF is fully
gapped in bulk and has three chiral gapless edge states located on the
boundary. More interestingly, the TSF can be modulated through initializing
the lasers. Furthermore, there is one stable Majorana fermion bound to each
vortex in the TSF, and the commensurability between the SF vortex lattice
structures and the TOLs shows advantages to investigate the properties of
Majorana fermions. Hence, these properties make the system a potential
candidate to perform QTC.

The paper is organized as follows. In Sec. II, we propose a scheme to simulate
the RSOC and ZF through the laser-atom interaction in triangular lattices and
an effective tight-binding Hamilton describing the fermionic atom in dark
states is deduced. In Sec. III, through applying the $s$-wave Feshbach
resonance, we discuss the properties of the TSF and NSF with mean-field
approximation. Furthermore, we discuss the Majorana zero mode in the vortex
structure in the TSF states. In Sec. IV, we summarize our results.

\section{\bigskip Simulation and Model Hamiltonian}

\begin{figure}[ptb]
\begin{center}
\includegraphics[width=1.0\linewidth]{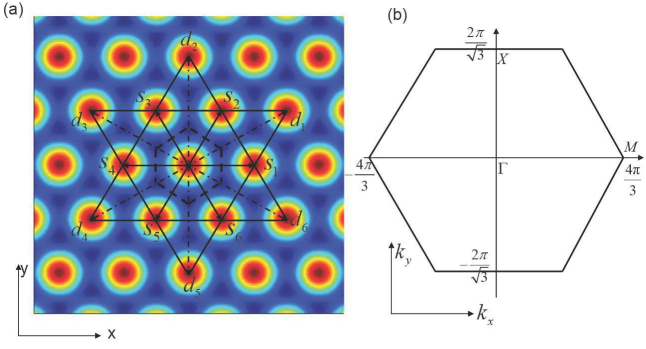}
\end{center}
\caption{(Color online) (a). Uniform triangle lattices are formed from the
maxima of the potential given by $V(\mathbf{r})=V_{L}\sum_{i=1}^{3}%
\cos(\mathbf{k}_{i}\cdot\mathbf{r})$. The defined lattice vectors
$\mathbf{s}_{i}$ and $\mathbf{d}_{i}$ are shown. The hexangular zone encircled
by the black-dashed lines is the unit cell of triangular lattice. (b) The
Brillouin zone (BZ) of triangular lattices, and the high-symmetry points are
marked.}%
\end{figure}

Firstly, we apply three blue detuned laser beams to create two-dimensional
TOLs, which can trap atoms at lattice sites. The three laser beams have same
wave-vector length but different polarizations and are applied along three
different directions: $\pm\frac{\sqrt{3}}{2}\hat{e}_{x}$ $-$ $\frac{1}{2}%
\hat{e}_{y}$ and $\hat{e}_{y}$, respectively. The total potential is given by
$V(\mathbf{r})=V_{L}\sum_{i=1}^{3}\cos(\mathbf{k}_{i}\cdot\mathbf{r})$ with
wave vectors $\mathbf{k}_{1,2}=k\left(  \pm\frac{\sqrt{3}}{2},\frac{1}%
{2}\right)  $ and $\mathbf{k}_{3}=k\left(  0,1\right)  $. The pattern of the
potential is shown in Fig.1(a), where the maxima form perfect TOLs.

In order to simulate the RSOC, we consider the ultra-cold fermionic atoms
trapped in the TOLs and having tripod-type level configuration (e.g., the
lowest three Zeeman levels of $^{6}$Li atoms near the broad $s$-wave Feshbach
resonance)\cite{Ruseckas,Stanescu,Zhu} shown in Fig. 2(a). Three degenerate
hyperfine ground states $|1\rangle$, $|2\rangle$ and $|3\rangle$ are coupled
to an excited state $|4\rangle$ through spatially modulated two sets of lasers
with the corresponding Rabi frequencies $\Omega_{v,1}$, $\Omega_{v,2}$ and
$\Omega_{v,3}$ with $v=a,b$ denoting two independent sets. The Rabi
frequencies can be parameterized as $\Omega_{v,1}=\frac{1}{2}\Omega_{v}%
\sin\theta_{v}\cos\phi_{v}e^{iS_{v,1}}$; $\Omega_{v,2}=\frac{1}{2}\Omega
_{v}\sin\theta_{v}\sin\phi_{v}e^{iS_{v,2}}$; $\Omega_{v,3}=\frac{1}{2}%
\Omega_{v}\cos\theta_{v}e^{iS_{v,3}}$, and $\sum_{i=1}^{3}\left\vert
\Omega_{v,i}\right\vert ^{2}=\frac{1}{4}\Omega_{v}^{2}$. Thus, the system can
be described by a Hamiltonian,%

\begin{equation}
H_{0}=H_{tb}+H_{l-a} \label{H_lat}%
\end{equation}
with%

\begin{equation}
H_{tb}=-\sum_{i,\alpha}\mu a_{i,\alpha}^{\dag}a_{i,\alpha}-\sum_{i,j,\alpha
,\beta}t_{i,j}a_{i,\alpha}^{\dag}a_{j,\beta} \label{H_tb}%
\end{equation}
and%

\begin{equation}
H_{l-a}=\underset{i,v}{\sum}\delta_{v}a_{i,4}^{\dag}a_{i,4}-\underset
{i,v,\alpha}{\sum}\hbar\Omega_{v,\alpha}a_{i,\alpha}^{\dag}a_{i,4}+H.c
\label{H_la}%
\end{equation}
Here, $H_{tb}$ is the tight-binding Hamiltonian describing the atom hopping
between different sites, and $H_{l-a}$ describes the laser-atom coupling.
$\mu$ is the chemical potential. $a_{i,\alpha}^{\dag}$ is the creation
operator of atom on site $i$ and in state $|\alpha\rangle$ with $\alpha$=1, 2,
3. $t_{ij}$ is the hopping integral between site $i$ and $j$. $\delta_{v}$ is
the detuning to the excited state $|4\rangle$.

Since the energy scale of $\delta_{v}$ and $\hbar\Omega_{v,\alpha}$ is much
larger than that of $\mu$ and $t_{i,j}$ (See the below discussions about the
parameters parts.), we firstly consider $H_{l-a}$. The eigenvalues of
$H_{l-a}$ can be obtained from the diagonalization. Namely, $E_{i,v,n=1,2,3,4}%
=0,0,\frac{1}{2}(\delta_{v}\mp\sqrt{\delta_{v}^{2}+\hbar^{2}\Omega_{v}^{2}})$.
The corresponding eigenstates (dressed states) are\begin{widetext}
\begin{align}
|D_{i,v,1}\rangle & =\sin\phi_{v}e^{iS_{v,3,1}}\left\vert 1\right\rangle
-\cos\phi_{v}e^{iS_{v,3,2}}\left\vert 2\right\rangle \nonumber\\
|D_{i,v,2}\rangle & =\cos\theta_{v}\cos\phi_{v}e^{iS_{v,3,1}}\left\vert
1\right\rangle +\cos\theta_{v}\sin\phi_{v}e^{iS_{v,3,2}}\left\vert
2\right\rangle -\sin\theta_{v}\left\vert 3\right\rangle \nonumber\\
|B_{i,v,1}\rangle & =\frac{1}{\sqrt{\Omega_{v}^{2}+(2E_{i,v,3})^{2}}}\left[
\Omega_{v}\left(  \sin\theta_{v}\cos\phi_{v}e^{iS_{v,3,1}}\left\vert
1\right\rangle +\sin\theta_{v}\sin\phi_{v}e^{iS_{v,3,2}}\left\vert
1\right\rangle +\cos\theta_{v}\left\vert 3\right\rangle \right)
-2E_{i,v,3}e^{iS_{v,3}}\left\vert 4\right\rangle \right]  \nonumber\\
|B_{i,v,2}\rangle & =\frac{1}{\sqrt{\Omega_{v}^{2}+(2E_{i,v,4})^{2}}}\left[
\Omega_{v}\left(  \sin\theta_{v}\cos\phi_{v}e^{iS_{v,3,1}}\left\vert
1\right\rangle +\sin\theta_{v}\sin\phi_{v}e^{iS_{v,3,2}}\left\vert
1\right\rangle +\cos\theta_{v}\left\vert 3\right\rangle \right)
-2E_{i,v,4}e^{iS_{v,3}}\left\vert 4\right\rangle \right].  \label{eigenstate}%
\end{align}
\end{widetext}Here, $|B_{i,v,1/2}\rangle$ represent zero-energy dark states,
while $|D_{i,v,1/2}\rangle$ represent the nonzero-energy bright states. That
means the energy of dark states is not adjusted by the laser fields. Moreover,
the dark states $|D_{i,v,1/2}\rangle$ have no coupling with the initial
excited state $|4\rangle$. Therefore, the dark states are stable under atomic
spontaneous emission. With the adiabatic approximation\cite{Zhu,Stanescu}, we
can neglect all the couplings that simultaneously involve the dark states and
bright states and reduce the Hamiltonian $H_{0}$ into the subspace spanned by
the dark states. In general, the dark states $|D_{i,a,1/2}\rangle$ produced by
laser set $a$ (See the red lines in Fig. 2 (a)) are different from the dark
states $|D_{i,b,1/2}\rangle$ produced by laser set $b$ (See the green lines in
Fig. 2 (a)). However, the two sets of dark states can be same through
initializing the parameters of the lasers, and the equivalence attributes to
the periodicity of TOLs. In present work, we concentrate on the two sets of
lasers configuration illustrated in Fig. 2 (b) and (c). The initialized
parameters for the laser fields are $\theta_{a}=k_{a,2}y+\varphi$, $\phi
_{a}=\frac{\pi}{4}$, $S_{a,1}=k_{a,1}x$, $S_{a,2}=-k_{a,1}x$, $S_{a,3}%
=k_{a,3}z$ and $\theta_{b}=k_{b,1}x+\varphi$, $\phi_{b}=k_{b,2}y+\frac{\pi}%
{4}$, $S_{b,1}=0$, $S_{b,2}=0$, $S_{b,3}=k_{b,3}z$, where $\varphi$ is an
arbitrary phase and $k_{v,1}$, $k_{v,2}$ and $k_{v,3}$ are the wave vectors of
the lasers along the $x$, $y$, $z$ axes, respectively. The wave vectors of the
lasers are initialized to fulfil the relations: $k_{v,1}$=$4\pi$, $k_{v,2}%
$=$4\pi/\sqrt{3}$, so that the commensuration with the TOLs is guaranteed
simultaneously. Now, $|D_{i,v,1}\rangle$=$\frac{\sqrt{2}}{2}(\left\vert
1\right\rangle -\left\vert 2\right\rangle )$, $|D_{i,v,2}\rangle$=$\frac
{\sqrt{2}}{2}(\left\vert 1\right\rangle +\left\vert 2\right\rangle
)\cos\varphi-\sin\varphi\left\vert 3\right\rangle $ for both $v=a$ and $b$.
That means the Hamiltonian (\ref{H_lat}) can be projected into the subspace
spanned by $|D_{i,\uparrow}\rangle$ and $|D_{i,\downarrow}\rangle$ with
$|D_{i,\uparrow}\rangle\equiv|D_{i,v,1}\rangle$ and $|D_{i,\downarrow}%
\rangle\equiv|D_{i,v,2}\rangle$ if the atoms are initially pumped to these
dark states and they remain in the dark states. Here, we use $\sigma
=\uparrow\downarrow$ to denote two pseudo-spin. Then, in the dark states
subspace, the Hamiltonian $H_{0}$ can be projected into the following form:

\begin{figure}[ptb]
\begin{center}
\includegraphics[width=1.0\linewidth]{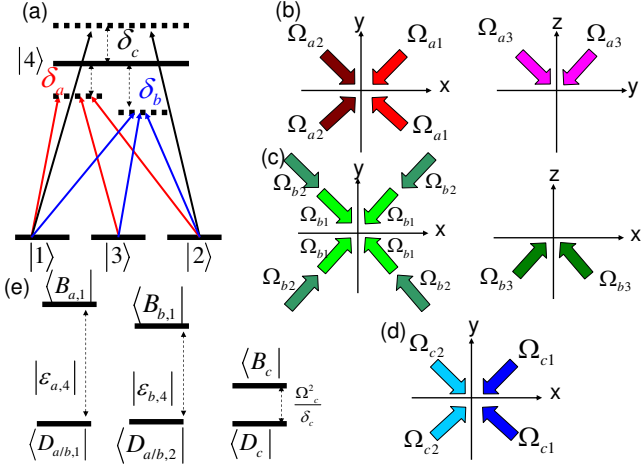}
\end{center}
\caption{(color online) Illustration of the light-atom interaction for
generation of effective non-Abelian gauge fields and effective Zeeman fields.
(a) The configuration of the hyperfine levels of ultra-cold atom and three
sets of laser beams characterized by the Rabi frequencies $\Omega_{a,i}$,
$\Omega_{b,i}$ and $\Omega_{c,j}$ with $i=1,2,3$ and $j=1,2$. The atom and the
laser fields have interaction through the Raman-type coupling with a large
single-photon detuning $\delta_{a/b/c}$. The laser beams configuration for
$\Omega_{a}$, $\Omega_{b}$ and $\Omega_{c}$ are shown in (b), (c) and (d). (e)
The relative energy levels modulated by the atom-laser couplings.}%
\end{figure}%

\begin{equation}
H_{eff}=-\sum_{i,\sigma}\mu c_{i,\sigma}^{\dag}c_{i,\sigma}-\sum
_{i,j,\sigma,\sigma^{\prime}}t_{i,j}c_{i,\sigma}^{\dag}U_{\sigma
,\sigma^{\prime}i,j}c_{j,\sigma^{\prime}}\text{.} \label{H_lat2}%
\end{equation}
Here, $c_{i,\sigma}^{\dag}$ is the creation operator of atom on site $i$ in
eigenstate $|D_{i,\uparrow}\rangle$. The Peierls phase factor is,%

\begin{equation}
U_{\sigma,\sigma^{\prime}i,j}=U_{\sigma,\sigma^{\prime}i}U_{\sigma
,\sigma^{\prime}j}^{\dag}=e^{i\int_{\mathbf{r}_{j}}^{\mathbf{r}_{i}%
}(\mathbf{\tilde{A}}_{a.\sigma,\sigma^{\prime}}+\mathbf{\tilde{A}}%
_{b,\sigma,\sigma^{\prime}})\cdot d\mathbf{r}} \label{phase factor}%
\end{equation}
Where $\mathbf{\tilde{A}}_{v.\sigma,\sigma^{\prime}}$ is the
laser-field-induced gauge vector potential, and $\mathbf{\tilde{A}}%
_{v,\sigma\sigma^{\prime}}$=$i\hbar\langle D_{v,\sigma}|\nabla|D_{v,\sigma
^{\prime}}\rangle$. We list the forms of $\mathbf{\tilde{A}}_{v,\sigma
\sigma^{\prime}}$ for completeness:%

\begin{align}
\mathbf{\tilde{A}}_{v,\uparrow\uparrow}  &  =\hbar\left(  \cos^{2}\phi
_{v}\nabla S_{v,2,3}+\sin^{2}\phi_{v}\nabla S_{v,1,3}\right) \nonumber\\
\mathbf{\tilde{A}}_{v,\uparrow\downarrow}  &  =\hbar\cos\theta_{v}\left(
\frac{1}{2}\sin2\phi_{v}\nabla S_{v,1,2}-i\nabla\phi_{v}\right) \nonumber\\
\mathbf{\tilde{A}}_{v,\downarrow\downarrow}  &  =\hbar\cos^{2}\theta
_{v}\left(  \cos^{2}\phi_{v}\nabla S_{v,1,3}+\sin^{2}\phi_{v}\nabla
S_{v,2,3}\right)  \label{vector_potential}%
\end{align}
Note that since the two sets of lasers $a$ and $b$ have different detunings
$\delta_{a}$ and $\delta_{b}$ to the excited state $|4\rangle$, there are no
interference effects between the two sets, and they interact with the atoms
independently. The total gauge vector potentials are the simple sum of
$\mathbf{\tilde{A}}_{a.\sigma,\sigma^{\prime}}+\mathbf{\tilde{A}}%
_{b,\sigma,\sigma^{\prime}}$.

Now, we evaluate that the effective RSOC can be simulated by the
aforementioned two sets of lasers $a$ and $b$. For convenience, we define the
(next) nearest neighbor lattice vectors: $\mathbf{s}_{n}$ ($\mathbf{d}_{n}$)
shown in Fig. 1 (a) with $n$=$1$...$6$, and set the lattice constant $1$. For
the two sets of lasers configuration illustrated in Fig. 2 (b) and (c), we can
find that $U_{\mathbf{s}_{1}}$=$U_{\mathbf{s}_{4}}^{\dag}$=$e^{i4\pi
\cos\varphi\sigma_{x}}$ and $U_{\mathbf{d}_{2}}$=$U_{\mathbf{d}_{5}}^{\dag}%
$=$e^{i4\pi\cos\varphi\sigma_{y}}$ are the only nontrivial phase factors and
other $U_{\mathbf{s}_{n}/\mathbf{d}_{n}}$ are trivial and equal $\mathbf{1}$.
Since the TOLs have the rotation symmetry of point group $C_{3\upsilon}$,
rotating the laser beams or the lattice systems with $\pm\frac{2\pi}{3}$ gives
another two groups of nontrivial phase factors: $U_{\mathbf{s}_{3}}%
$=$U_{\mathbf{s}_{6}}^{\dag}$=$e^{i4\pi\cos\varphi\sigma_{x}}$, $U_{\mathbf{d}%
_{1}}$=$U_{\mathbf{d}_{4}}^{\dag}$=$e^{-i4\pi\cos\varphi\sigma_{y}}$ and
$U_{\mathbf{s}_{2}}$=$U_{\mathbf{s}_{5}}^{\dag}$=$e^{-i4\pi\cos\varphi
\sigma_{x}}$, $U_{\mathbf{d}_{3}}$=$U_{\mathbf{d}_{6}}^{\dag}$=$e^{-i4\pi
\cos\varphi\sigma_{y}\text{ }}$, respectively. Here, $\sigma_{x/y}$ are the
two Pauli matrix. For a two-component spin system, we have the following
relation for the unitary operator:%

\begin{equation}
U_{\sigma\sigma^{\prime}}=e^{i\alpha(\sigma_{x/y})_{\sigma\sigma^{\prime}}%
}=\cos\alpha+i(\sigma_{x/y})_{\sigma\sigma^{\prime}}\sin\alpha. \label{U1}%
\end{equation}
With Eq. (\ref{U1}), we can find that $H_{eff}$ in Eq. (\ref{H_lat2}) has the
form as follows, \begin{widetext}%
\begin{align}
H_{eff}  & =-\sum_{i,\sigma}\mu c_{i,\sigma}^{\dag}c_{i,\sigma}-t\cos
\alpha\sum_{i,,n,\sigma}c_{i,\sigma}^{\dag}c_{i+\mathbf{s}_{n},\sigma
}-t^{\prime}\cos\alpha\sum_{i,,n,\sigma}c_{i,\sigma}^{\dag}c_{i+\mathbf{d}%
_{n},\sigma}\nonumber\\
& -it\sin\alpha\sum_{i,,n,\sigma,\sigma^{\prime}}(-1)^{n+1}c_{i,\sigma}^{\dag
}(\sigma_{x})_{\sigma\sigma^{\prime}}c_{i+\mathbf{s}_{n},\sigma^{\prime}%
}-it^{\prime}\sin\alpha\sum_{i,,n,\sigma,\sigma^{\prime}}(-1)^{n}c_{i,\sigma
}^{\dag}(\sigma_{y})_{\sigma\sigma^{\prime}}c_{i+\mathbf{d}_{n},\sigma
^{\prime}}.\label{H_effn}%
\end{align}
\end{widetext} Here, $t$ and $t^{\prime}$ are the original nearest and
next-nearest neighbor hopping integrals. $\alpha=4\pi\cos\varphi$. The first
three terms in Eq. (\ref{H_effn}) are the modulated normal hopping parts of
the Hamiltonian, while the last two terms describe the effective RSOC. More
importantly, through adjusting the gauge flux $\alpha$, one can change the
relative strength between the hopping and RSOC. That is nearly impossible in
the condensed matter system. The $k^{3}$ type of RSOC can be explicitly found
in the momentum space form of Eq. (\ref{H_effn}), which we will discuss in the
next section.

In the following part of this section, we simulate how to generate an
effective ZF to split two pseudo-spin states $|D_{i,\uparrow}\rangle$ and
$|D_{i,\downarrow}\rangle$. We apply two additional laser beams that couple
the states $\left\vert 1\right\rangle $ and $\left\vert 2\right\rangle $ to
the excited state $\left\vert 4\right\rangle $ with a large detuning
$\delta_{c}$\cite{Zhu1} (See the blue lines in Fig. 2 (a) ). The laser-atom
interaction is
\begin{equation}
H_{l-a}^{\prime}=-\underset{i}{\sum}\delta_{c}a_{i,4}^{\dag}a_{i,4}%
+\underset{i,\alpha=1}{\overset{\alpha=2}{\sum}}\hbar\Omega_{c,\alpha
}a_{i,\alpha}^{\dag}a_{i,4}+H.c. \label{H_lan}%
\end{equation}
The corresponding Rabi frequencies are parameterized as $\Omega_{c,1}%
=\frac{\sqrt{2}}{4}\Omega_{c}e^{i4\pi x}$ and $\Omega_{c,2}=\frac{\sqrt{2}}%
{4}\Omega_{c}e^{-i4\pi x}$ with $\sqrt{\left\vert \Omega_{c,1}\right\vert
^{2}+\left\vert \Omega_{c,2}\right\vert ^{2}}=\frac{1}{2}\Omega_{c}$ and
$\hbar\Omega_{c}\ll$ $\delta_{c}$. The lasers configuration is illustrated in
Fig. 2 (d).

The eigenvalues of $H_{l-a}^{\prime}$ can be obtained from the
diagonalization. Namely, $E_{i,n=1,2,3}^{\prime}=0,-\frac{1}{2}(\delta_{c}%
\mp\sqrt{\delta_{c}^{2}+\hbar^{2}\Omega_{c}^{2}})$. The corresponding
eigenstates are:%

\begin{align}
\left\vert \chi_{1}\right\rangle  &  =\frac{\sqrt{2}}{2}e^{-i4\pi x}\left\vert
1\right\rangle -\frac{\sqrt{2}}{2}e^{i4\pi x}\left\vert 2\right\rangle
\nonumber\\
\left\vert \chi_{2}\right\rangle  &  =\frac{\sqrt{2}}{2}\cos\beta e^{-i4\pi
x}\left\vert 1\right\rangle +\frac{\sqrt{2}}{2}\cos\beta e^{i4\pi x}\left\vert
2\right\rangle -\sin\beta\left\vert 4\right\rangle \nonumber\\
\left\vert \chi_{3}\right\rangle  &  =\frac{\sqrt{2}}{2}\sin\beta e^{-i4\pi
x}\left\vert 1\right\rangle +\frac{\sqrt{2}}{2}\sin\beta e^{i4\pi x}\left\vert
2\right\rangle +\cos\beta\left\vert 4\right\rangle \label{eigenstate2}%
\end{align}
Here, $\tan\beta=(\sqrt{\delta_{c}^{2}+\hbar^{2}\Omega_{c}^{2}}-\delta
_{c})/\delta_{c}$. Due to $\hbar\Omega_{c}\ll$ $\delta_{c}$, we can get
$\tan\beta\sim\hbar\Omega_{c}/\delta_{c}\sim0$, and $E_{i,2}^{\prime}\sim
\hbar^{2}\Omega_{c}^{2}/4\delta_{c}$.
\begin{equation}
\left\vert \chi_{2}\right\rangle \sim\frac{\sqrt{2}}{2}e^{-i4\pi x}\left\vert
1\right\rangle +\frac{\sqrt{2}}{2}e^{i4\pi x}\left\vert 2\right\rangle
,\left\vert \chi_{3}\right\rangle \sim\left\vert 4\right\rangle . \label{chi3}%
\end{equation}
Since $E_{i,1}^{\prime}=0$, there is no effect of $\left\vert \chi
_{1}\right\rangle $ to the ground states $\left\vert 1\right\rangle $ and
$\left\vert 2\right\rangle $, and $\left\vert \chi_{4}\right\rangle $ also has
no effect to $\left\vert 1\right\rangle $ and $\left\vert 2\right\rangle $.
Hence, we can only consider the effect of $\left\vert \chi_{2}\right\rangle $
to $\left\vert 1\right\rangle $ and $\left\vert 2\right\rangle $. If we define
that $d_{i}^{\dag}$ is an operator to create an atom on site $i$ in eigenstate
$\left\vert \chi_{2}\right\rangle $. A perturbation Hamiltonian can be written
as: $H_{p}=$ $\hbar\Omega_{p}\sum_{i}d_{i}^{\dag}d_{i}$. With Eq.
(\ref{chi3}), $H_{p}$ has the form:%

\begin{equation}
H_{p}=H_{ac}+\hbar\Omega_{p}\underset{i}{\sum}e^{-i8\pi x}a_{i,1}^{\dag
}a_{i,2}+H.c. \label{Hp2}%
\end{equation}
Where $\Omega_{p}=$ $\hbar\Omega_{c}^{2}/8\delta_{c}$. $H_{ac}=$ $\hbar
\Omega_{p}\underset{i}{\sum}(a_{i,1}^{\dag}a_{i,1}+a_{i,2}^{\dag}a_{i,2})$ is
a constant ac-Stark shift, whose effect can be canceled with a frequency
offset of the laser beams $\Omega_{v,3}$ applied to the level $\left\vert
3\right\rangle $\cite{Zhu1}. Therefore, we can only consider the effect of the
second term in Eq. (\ref{Hp2}). From Eq. (\ref{eigenstate}), we can get the
following relations%

\begin{align}
\left\vert 1\right\rangle  &  =\frac{\sqrt{2}}{2}\left(  \left\vert
D_{i,\uparrow}\right\rangle +\cos\varphi\left\vert D_{i,\downarrow
}\right\rangle \right) \nonumber\\
\left\vert 2\right\rangle  &  =\frac{\sqrt{2}}{2}\left(  -\left\vert
D_{i,\uparrow}\right\rangle +\cos\varphi\left\vert D_{i,\downarrow
}\right\rangle \right)  \label{eigen3}%
\end{align}
Where we have applied the conditions that at all the lattice sites, $e^{\pm
i8\pi x}=1$, $e^{\pm iS_{v,1,2}}=1$, $\theta=\varphi$ and $\phi=\pi/4$. With
Eq. (\ref{eigen3}), we find the second term in Eq. (\ref{Hp2}) induce a
splitting between $|D_{i,\uparrow}\rangle$ and $|D_{i,\downarrow}\rangle$ as:%

\begin{align*}
H_{s}  &  =-\hbar\Omega_{p}\underset{i}{\sum}(c_{i,\uparrow}^{\dag
}c_{i,\uparrow}-\cos^{2}\varphi c_{i,\downarrow}^{\dag}c_{i,\downarrow})\\
&  =-h_{0}\underset{i,\sigma}{\sum}c_{i,\sigma}^{\dag}c_{i,\sigma}%
-h_{z}\underset{i}{\sum}(c_{i,\uparrow}^{\dag}c_{i,\uparrow}-c_{i,\downarrow
}^{\dag}c_{i,\downarrow}),
\end{align*}
with $h_{0}=\hbar\Omega_{p}(1-\cos^{2}\varphi)/2$ and $h_{z}=\hbar\Omega
_{p}(1+\cos^{2}\varphi)/2$. $h_{0}$ can be renormalized into the chemical
potential term in $H_{eff}$ (Eq. (\ref{H_effn})), and $h_{z}$ describes the
effective ZF. In order to guarantee that $H_{s}$ cannot pump the atoms outside
of the dark-state subspace, the conditions: $\hbar\Omega_{p}\ll\left\vert
E_{b,3}\right\vert $ $<\left\vert E_{a,3}\right\vert $ must be fulfilled (See
Fig. 2 (e)). Now, the new Hamiltonian including the effective RSOC and ZF is:%

\begin{equation}
H_{0}^{\prime}=H_{eff}+H_{s}. \label{H_lat3}%
\end{equation}

\section{Topological SF and Majorana Fermion}

The SF states can be induced by atomic interaction from the s-wave scattering.
The interaction term is described by the Hamiltonian:%

\begin{equation}
H_{int}=\sum_{i}\sum_{\alpha<\beta}V_{\alpha\beta}a_{i,\alpha}^{\dag
}a_{i,\beta}^{\dag}a_{i,\beta}a_{i,\alpha}, \label{H_int}%
\end{equation}
where $\alpha$ and $\beta$ label three ground states of atoms and
$V_{\alpha\beta}$ are proportional to $s$-wave scattering lengths between
$\alpha$, $\beta$ channel. In $s$-wave SF state, $H_{int}$ can be decoupled on
the mean-field level:%

\begin{equation}
H_{mf}=\sum_{i}\sum_{\alpha<\beta}\Delta_{\alpha\beta}a_{i,\alpha}^{\dag
}a_{i,\beta}^{\dag}+H.c. \label{H_int1}%
\end{equation}
with $\Delta_{\alpha\beta}=V_{\alpha\beta}\left\langle a_{i,\beta}a_{i,\alpha
}\right\rangle $, the SF order parameter. Under the condition of
$\Delta_{\alpha\beta}\ll\Omega_{a/b}$, it is safe to consider the SF in the
dark-state subspace, because $H_{mf}$ cannot pump the atoms outside of the
dark-state subspace. Then, we project $H_{mf}$ to the dark-state subspace and have%

\begin{equation}
H_{mf}^{\prime}=\sum_{i}\Delta_{0}c_{i,\uparrow}^{\dag}c_{i,\downarrow}^{\dag
}+H.c. \label{H_int2}%
\end{equation}
with $\Delta_{0}$ the linear combinations of $\Delta_{\alpha\beta}$.

In the following parts of the paper, we focus on the total Hamiltonian which
describes the SF states:%

\begin{equation}
H_{t}=H_{0}^{\prime}+H_{mf}^{\prime}\text{.} \label{H_tot2}%
\end{equation}
After the Fourier transformation, in momentum Nambu bases:
$[c_{\mathbf{k\uparrow}},c_{\mathbf{k}\downarrow,}c_{-\mathbf{k\uparrow}%
}^{\dag},c_{-\mathbf{k\downarrow}}^{\dag},]^{T}$, $H_{t}$ can be expressed as:%

\begin{equation}
H_{t}(\mathbf{k})=\left[
\begin{array}
[c]{cc}%
\varepsilon_{\mathbf{k}}-h_{z}\sigma_{z}+\mathbf{g}_{\mathbf{k}}%
\cdot\mathbf{\sigma} & -i\Delta_{0}\sigma_{y}\\
i\Delta_{0}\sigma_{y} & -(\varepsilon_{\mathbf{k}}-h_{z}\sigma_{z}%
)+\mathbf{g}_{\mathbf{k}}\cdot\mathbf{\sigma}^{\ast}%
\end{array}
\right]  . \label{H_T}%
\end{equation}
Here $\mathbf{g}_{\mathbf{k}}=(a_{\mathbf{k}},b_{\mathbf{k}})$,
$\mathbf{\sigma}=(\sigma_{x},\sigma_{y})$, and the explicit forms of
$\varepsilon_{\mathbf{k}}$, $a_{\mathbf{k}}$ and $b_{\mathbf{k}}$ are listed:%

\begin{align}
\varepsilon_{\mathbf{k}}  &  =-2t_{1}(\cos k_{x}+2\cos\frac{k_{x}}{2}\cos
\frac{\sqrt{3}k_{y}}{2})\nonumber\\
&  -2t_{3}(\cos\sqrt{3}k_{y}+2\cos\frac{3k_{x}}{2}\cos\frac{\sqrt{3}k_{y}}%
{2})-\mu^{\prime} \label{Ekk}%
\end{align}

\begin{align}
a_{\mathbf{k}}  &  =2t_{2}(\sin k_{x}-2\sin\frac{k_{x}}{2}\cos\frac{\sqrt
{3}k_{y}}{2})\nonumber\\
b_{\mathbf{k}}  &  =2t_{4}(-\sin\sqrt{3}k_{y}+2\sin\frac{\sqrt{3}k_{y}}{2}%
\cos\frac{3k_{x}}{2}) \label{akbk}%
\end{align}
in which $t_{1}$=$t\cos\alpha$, $t_{2}$=$t\sin\alpha$, $t_{3}$=$t^{\prime}%
\cos\alpha$, $t_{4}$=$t^{\prime}\sin\alpha$ with $\alpha=4\pi\cos\varphi$ and
$\mu^{\prime}=\mu+h_{0}$.

Before discussing the properties of the SF states described by $H_{t}$ in Eq.
(\ref{H_tot2}), we give the estimations about the parameters related to the
aforementioned simulations to ensure the experimental feasibility and
rationality. For trapped atoms: $^{6}$Li, the wave length of laser beams
utilized to produce the TOL is $\lambda_{L}\sim1\mu m$, and the lattice
constant is $a_{L}=\frac{2\lambda_{L}}{\sqrt{3}}$. The recoil energy is
$E_{r}=\frac{\hbar^{2}k_{L}^{2}}{2m}\sim\hbar\times2\pi\times30$ kHZ$\sim1\mu
K$ with $k_{L}=\frac{2\pi}{\lambda_{L}}$, and $t=\frac{4E_{r}}{\sqrt{\pi}%
}(\frac{V_{0}}{E_{r}})^{\frac{3}{4}}e^{-2\sqrt{V_{0}/E_{r}}}$\cite{Zwerger}
when $V_{0}>>$ $E_{r}$ with $V_{0}\sim4V_{L}$, the depth of the TOL.
$t^{\prime}/t=e^{-\eta(\sqrt{3}-1)\sqrt{(V_{0}-E_{kin})/E_{r}}k_{L}a_{L}}$
with $E_{kin}$ and $\eta$ the kinetic energy of atom and the renormalized
factor. The typical atomic velocity is about several centimeters per second,
and $E_{kin}$ has the same order of $E_{r}$. $\eta$ depends on geometry of the
lattice. According to the 1D lattice results\cite{Bloch} and $\eta<1$, we
estimate that $V_{0}\sim3E_{r}$ is enough to get $t^{\prime}/t=\frac{1}%
{3\sqrt{3}}$. Actually, the TSF is robust even when $t^{\prime}/t\sim10^{-3}$.
Here, without lost of generality, we set $t^{\prime}/t\equiv\frac{1}{3\sqrt
{3}}$. Then $t\sim0.1E_{r}$ with the aforementioned formula. From the
harmonic-potential approximation, the energies of the atoms tightly confined
at a single lattice site are quantized to levels separated by $2\pi\hbar
\omega_{0}$=$2E_{r}\sqrt{\frac{V_{0}}{E_{r}}}$\cite{Bloch}. On the other hand,
the maximal band width for the TOL is $W_{m}\sim10t$ $\sim E_{r}$. Therefore,
it is safe to describe the system with single-band approximation because of
$2\pi\hbar\omega_{0}>W_{m}$. The Rabi frequencies $\Omega_{a/b/c}$ are
$10^{3}E_{r}/\hbar$, and the $\Omega_{p}$ can be tuned from $0$ to
$E_{r}/\hbar$ which is enough for $\hbar\Omega_{p}\ll\left\vert E_{b,3}%
\right\vert $ $<\left\vert E_{a,3}\right\vert $. Then, adiabatic
approximation\cite{Zhu1} is reasonable. The typical $s$-wave pairing potential
$\Delta_{\alpha\beta}$ in experiments is about $0.1E_{r}/\hbar$\cite{Ketterle}%
, which is much smaller than $\Omega_{a/b}$. Hence, our proposal is
experimentally feasible when the parameters lie in the estimated region.

\bigskip For convenience to discuss the properties of SF, we rewrite
$H_{t}(\mathbf{k})$ in the new bases $[\hat{\psi}_{\mathbf{k,}+},\hat{\psi
}_{-\mathbf{k,}+}^{\dag},-\hat{\psi}_{-\mathbf{k,}-}^{\dag},-\hat{\psi
}_{\mathbf{k,}-}]^{T}$ with $\hat{\psi}_{\mathbf{k,}\pm}$= $\frac{1}{\sqrt{2}%
}\left[  c_{\mathbf{k\uparrow}}\pm c_{-\mathbf{k\downarrow}}^{\dag}\right]  $,
$H_{t}^{\prime}$ has the following form:%

\begin{equation}
H_{t}^{\prime}(\mathbf{k})=\left[
\begin{array}
[c]{cc}%
H_{+}(\mathbf{k}) & -i\varepsilon_{\mathbf{k}}\sigma_{y}\\
i\varepsilon_{\mathbf{k}}\sigma_{y} & H_{-}(\mathbf{k})
\end{array}
\right]  . \label{H_R}%
\end{equation}
Here,%

\begin{equation}
H_{\pm}(\mathbf{k})=\pm\left[  (-h_{z}\mp\Delta_{0})\sigma_{z}+a_{k}\sigma
_{x}\pm b_{k}\sigma_{y}\right]  . \label{Hpm0}%
\end{equation}
The spectrums of Hamiltonian(\ref{H_R}) are $\pm E_{\pm}(k)$, and%

\begin{equation}
E_{\pm}(k)=\sqrt{\varepsilon_{\mathbf{k}}^{2}+\Delta_{0}^{2}+\Theta_{k}^{2}%
\pm2\sqrt{h_{z}^{2}\Delta_{0}^{2}+\varepsilon_{\mathbf{k}}^{2}\Theta_{k}^{2}}%
}, \label{Epm}%
\end{equation}
in which $\Theta_{k}=\sqrt{h_{z}^{2}+\left\vert \mathbf{g}_{\mathbf{k}%
}\right\vert ^{2}}$. The topological transition point is determined by bulk
energy gap closing condition: $E_{-}(k)=0$ (Fig. 3(d)). The spectrums are
fully gapped off this point, and the SF is topologically nontrivial when
$h_{z}>\sqrt{\Delta_{0}^{2}+\varepsilon_{\mathbf{k}}^{2}}|_{\mathbf{k}=(0,0)}$
(Fig. 3(c)) and trivial when $h_{z}<\sqrt{\Delta_{0}^{2}+\varepsilon
_{\mathbf{k}}^{2}}|_{\mathbf{k}=(0,0)}$ (Fig. 3(e)). Around the $\Gamma$ point
in BZ,
\begin{equation}
\mathcal{H}_{\pm}(\mathbf{k})=\mp\lbrack(\Delta_{0}\pm h_{z})\sigma
_{z}+\lambda_{so}(k_{-}^{3}\sigma_{\pm}+k_{+}^{3}\sigma_{\mp})]\text{,}
\label{H_plusandminus}%
\end{equation}
where $\sigma_{\pm}$=$\frac{1}{2}\left(  \sigma_{x}\pm\sigma_{y}\right)  $,
$k_{\pm}$=$\frac{1}{2}\left(  k_{x}\pm k_{y}\right)  $. The term $\lambda
_{so}(k_{-}^{3}\sigma_{\pm}+k_{+}^{3}\sigma_{\mp})$ is the $k^{3}$ RSOC with
amplitude: $\lambda_{so}$=$\frac{t_{2}}{2}\sim-0.48t$ for $\cos\varphi
=-0.101$. It is explicit that $\mathcal{H}_{\pm}(\mathbf{k})$ has the
well-defined $f$-wave chirality. That means the topological Chern
number\cite{Thouless} can be calculated $\mathcal{C}_{\mathcal{H}_{-}}$%
=$\frac{3}{2}\left[  \text{sign}(h_{z}-\Delta_{0})_{h_{z}>\Delta_{0}%
}-\text{sign}(h_{z}-\Delta_{0})_{h_{z}<\Delta_{0}}\right]  $=$3$ while
$\mathcal{C}_{\mathcal{H}+}$=$0$. From the square lattice results that
$\Delta_{0}$ has maximum when the filling $\sim$ 1 atom per
site\cite{Hofstetter}, we set $\Delta_{0}=0.5t$ and $\hbar\Omega_{p}%
=3\Delta_{0}$ for the chemical potential around $\mu_{1}$ and filling about
0.6 atom per site. When the chemical potential locates at the region of
$\mu_{4}$ (Fig. 3 (b)), the low-energy behaviors are dominated by
$H_{t}^{\prime}(\mathbf{k})$ with $\mathbf{k}\sim(0,\frac{2\sqrt{3}\pi}{3})$.
Then we cannot define the specific chirality, and we ascribe these cases to
NSF. According to the aforementioned analysis, we draw the phase diagram in
Fig. 3(e). We find that the TSF strongly depends on the initial parameter
$\varphi$ of the laser beams and the fillings. That means we can control the
topological properties of the SF by modulating the parameters of the lasers.
That provides convenience to investigate the TSF.

In lattice case, the ground state Chern number of $H_{t}$ can be calculated
with:
\begin{equation}
\mathcal{C}_{n}=\frac{1}{2\pi}\int_{BZ}d^{2}k2Im\langle\frac{\partial
u_{n}(\mathbf{k})}{\partial k_{x}}|\frac{\partial u_{n}(\mathbf{k})}{\partial
k_{y}}\rangle, \label{chernnumber1}%
\end{equation}
where $u_{n}(\mathbf{k})$ is the ground state wave-function for the $n$th
occupied band ($n$=1,2). The straightforward calculation gives $\mathcal{C}%
_{1}$=$0$ and $\mathcal{C}_{2}$=$3$ when $h_{z}>\sqrt{\Delta_{0}%
^{2}+\varepsilon_{\mathbf{k}}^{2}}|_{\mathbf{k}=(0,0)}$, and $\mathcal{C}_{1}%
$=$0$ and $\mathcal{C}_{2}$=$0$ when $h_{z}<\sqrt{\Delta_{0}^{2}%
+\varepsilon_{\mathbf{k}}^{2}}|_{\mathbf{k}=(0,0)}$. The non-zero Chern number
means the same numbers of gapless edge states from the bulk-edge
correspondence. From the energy spectrum of $H_{t}(k_{y},x)$ shown in Fig. 4,
we find that three gapless chiral edge states transport on one edge of TSF.
(see Fig.4 (a) and (d)). The effective Hamiltonian describing the chiral edge
states is
\begin{equation}
\mathcal{H}_{edge}=\sum_{k_{y}\geq0}\nu_{0}k_{y}\hat{\Psi}_{k_{y}}^{\dag
}(x)\hat{\Psi}_{k_{y}}(x), \label{Hedge1}%
\end{equation}
where $\nu_{0}$ is the effective velocity at $\Gamma$, and $\hat{\Psi}_{k_{y}%
}^{\dag}(x)=\sum_{\sigma}\int dx\left[  u_{k_{y},\sigma}(x)c_{\sigma}^{\dag
}(x)+v_{k_{y},\sigma}(x)c_{\sigma}(x)\right]  $. The Majorana condition
requires $\hat{\Psi}_{-k_{y}}(x)=$ $\hat{\Psi}_{k_{y}}^{\dag}(x)$, i.e.,
$u_{-k_{y,\sigma}}(x)=v_{k_{y,\sigma}}^{\ast}(x)$. We check that it's indeed
the case in our model. That means three edge states are chiral Majorana edge
states. \begin{figure}[ptb]
\begin{center}
\includegraphics[width=1.0\linewidth]{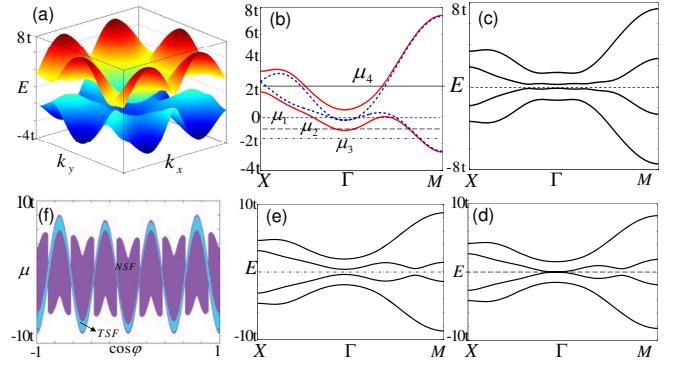}
\end{center}
\caption{(color online) (a) The band structures of $E_{k}=\varepsilon_{k}%
\pm\Theta_{k}$ with $\cos\varphi=-0.101$ and $\mu=-2.59$. Here, we set
$k_{x}\in\lbrack-\frac{4\pi}{3},\frac{4\pi}{3}]$ and $k_{y}\in\lbrack
-\frac{2\pi}{\sqrt{3}},\frac{2\pi}{\sqrt{3}}]$. (b) The band structures along
high symmetric lines (Fig.2 (b)). The dashed blue lines are $\varepsilon
_{k}\pm\left\vert g_{k}\right\vert $ and the solid red lines correspond to
(a). Four different fillings with chemical potential $\mu_{1/2/3/4}=-2.59t$,
$-3.436t$, $-3.89t$, $0.45t$ are shown. (c) (d) (e) are the SF quasi-particle
spectrums corresponding to $\mu_{1}$, $\mu_{2}$ and $\mu_{3}$. (f) The phase
diagram as change of $\cos\varphi$ and $\mu$. Two different phases, NSF and
TSF are identified. $\Delta_{0}=0.5t$ and $\hbar\Omega_{p}=3\Delta_{0}$. }%
\end{figure}\begin{figure}[ptbptb]
\begin{center}
\includegraphics[width=1.0\linewidth]{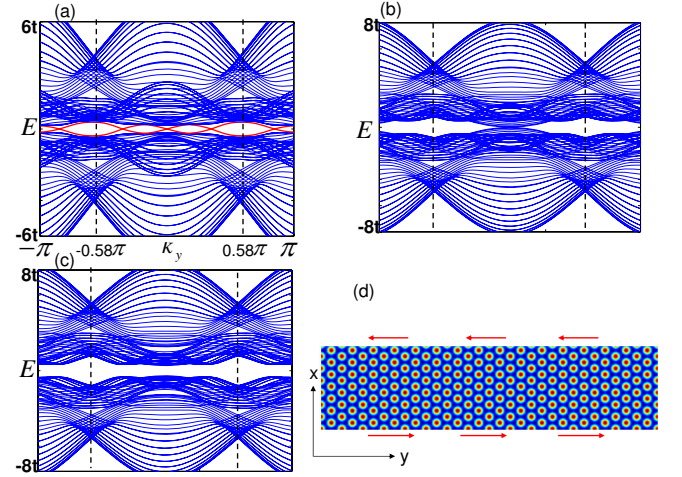}
\end{center}
\caption{(color online) The spectrums of Hamiltonian (\ref{H_tot2}) with edges
at x direction $i_{x}$ $\in(1,31)$. (a) (b) and (c) correspond to $\mu_{1}$
$\mu_{2}$ and $\mu_{3}$ cases in Fig. 3 (b). The dashed black lines indicate
the contributions from the first BZ. (d) The edge states (the states crossing
the gap and denoted with the solid red lines in (a)) transport along two
edges. }%
\end{figure}

In general, the topological defects bound Majorana zero modes in
TSC/TSF\cite{Qi,Read}. Here, we consider the topological excitations of vortex
structures in our system. The quasi-particle excitations usually are described
by the Bogoliubov-de Gennes (BdG) equation. From Fig. 4(a), we can find that
the wave function of low energy excitations can be constructed from the
contributions of quasi-particle around $\Gamma$. For simplicity, only the
non-trivial part $\mathcal{H}_{-}(k)$ of Eq.(\ref{H_plusandminus}) is taken
into account. Define $z=x+iy$, then $\partial_{z/z^{\ast}}=\partial_{x}\pm
i\partial_{y}$. The BdG equation for the quasi-particle has the form:
$\mathcal{H}_{-}(z,z^{\ast})\Psi_{0}=E\Psi_{0}$ with $\Psi_{0}=[u_{0}^{\ast
},v_{0}^{\ast}]^{T}$ and the corresponding quasi-particle creation operator:
$\hat{\Psi}_{0}^{\dag}=%
{\displaystyle\int}
dzdz^{\ast}\left[  u_{0}\hat{\psi}_{-}^{\dag}+v_{0}\hat{\psi}_{-}\right]  $.
In the uniform TSF states, we can assume a trivial wave function:
$u_{0}=e^{i\pi/4}z^{-\frac{3}{2}}e^{-\frac{2}{3}(\frac{h_{z}-\Delta_{0}%
}{\lambda_{so}})^{1/3}(zz^{\ast})^{\frac{3}{2}}}$ and $v_{0}=u_{0}^{\ast}$ as
a test wave function from the BdG equation. The SF order parameter with a
vortex structure can be approximately expressed as $\Delta_{0}(r)$=$0$ for
$r<r_{c}$ and $\Delta_{0}(r)$=$\Delta_{0}e^{i\theta}$ for $r>r_{c}$ with
vorticity $1$. We imagine that the vortex is created adiabatically by changing
the wave function slowly enough so that it always remains an eigenstate. The
wave function of the vortex state can be obtained from a singular gauge
transformation: $\Psi_{0}\rightarrow\Psi_{0}e^{iq\theta}$ with $q=\pm1$
identifying the quasi-particle and quasi-anti-particle. In reverse, the vortex
can be gauged away by the inverse singular gauge transformation: $k\rightarrow
k-\nabla\theta/2$ and $\Delta_{0}e^{i\theta}\rightarrow\Delta_{0}$
\cite{Sato2}. After the inverse transformation, the state is one of
eigenstates, namely, the vortex excited state. In analogy to the Laughlin's
argument\cite{Laughlin} about the vortex excitation in quantum Hall state, we
get the wave function describing vortex zero mode as $u_{0}^{\prime}\sim
e^{-i(\frac{\theta}{2}-\frac{\pi}{4})}r^{-\frac{3}{2}}e^{-\frac{2}{3}\left(
\frac{h_{z}-\Delta_{0}}{\lambda_{so}}\right)  ^{\frac{1}{3}}r^{3}}$ and
$v_{0}^{\prime}=(u_{0}^{\prime})^{\ast}$. The unique one zero mode for
$f$-wave case is proven by the numerical calculation\cite{Mao}.

The stability of Majorana zero mode is measured by the mini-gap $E_{g}%
\sim\Delta_{0}^{2}/E_{f}$ with $E_{f}$ the Fermi energy. In our case,
the\ $E_{f}$ is measured by $t$ not $E_{kin}$. Hence, The ratio $E_{g}%
/\Delta_{0}$ can be large enough to protect Majorana zero mode. Take half
filling as an example, we assume the optimized $\Delta_{0}\sim t$ and roughly
estimate $E_{f}\sim$ $3t$. Then $E_{g}/\Delta_{0}\sim1/3$. Comparing with the
hybridized systems \cite{Fu,Sau,Mao}, the energy scale of $E_{f}$ has the
order of electrons' kinetic energy $E_{kin}$ and $\Delta_{0}$ from the
proximity effect is much smaller compared to $E_{kin}$. So, the mini-gap in
hybridized systems may be relative small compare to superconductive gap
$\Delta_{0}$.

\section{\bigskip Conclusions}

In summary, we have proposed a scheme to produce $k^{3}$ RSOC and ZF through
the laser-atom interaction in TOLs, and a novelly chiral $f$-wave TSF is
realized thanks to the $s$-wave Feshbach resonance. We find that there exists
three Majorana edge states locating on the boundary of the system and one
Majorana fermion bounding to each vortex in the TSF state. The TSF can be
controlled by modulating the parameters of the laser. The controllability
provides convenience to investigate the properties of the TSF. Our proposal
enlarges TSF family and presents some advantages to study the Majorana fermions.

\emph{Acknowledgments:} Ningning Hao thanks J. Li and Guocai Liu thanks S. L.
Zhu for helpful discussions. The work is supported by the Ministry of Science
and Technology of China 973 program(2012CB821400), NSFC-1190024, NSFC-11147171
and NSFC-11247011.

\end{document}